\documentclass{ws-p9-75x6-50}

\def\be{\begin{equation}}
\def\ee{\end{equation}}
\def\ba{\begin{eqnarray}}
\def\ea{\end{eqnarray}}

\begin{document}

\title{Blue spectral inflation}

\author{Franz E. Schunck$^1$ and Eckehard W. Mielke$^2$}

\address{
$^1$Institut f\"ur Theoretische Physik,
Universit\"at zu K\"oln, 50923 K\"oln, Germany\\E-mail: fs@thp.uni-koeln.de\\
$^2$Departamento de F\'{\i}sica,
Universidad Aut\'onoma Metropolitana--Iztapalapa,
Apartado Postal 55-534, C.P. 09340, M\'exico, D.F., MEXICO\\
E-mail: ekke@xanum.uam.mx
}

\maketitle

The next-to leading order in the slow-roll approximation  of {\em inflation}
is known\cite{SM00} to lead to  the nonlinear  equation
\be
2 C \epsilon y^2 {\epsilon}''  -
(3\epsilon +1) y {\epsilon}'
+ \epsilon^2  + \epsilon + \Delta  = 0  \; ,
\label{Abel}
\ee
 of Stewart and Lyth\cite{stelyt}, where  the energy density $\epsilon=\epsilon(y)$ depends on  $y=H^2$,
the squared Hubble expansion rate, $C:=-2+\ln 2 +\gamma \simeq -0.73$, and $\Delta:= (n_{s}-1)/2$
denotes the deviation from the
scale invariant Harrison--Zel`dovich spectrum.

It is gratifying that
some of our new  {\em discrete eigenvalues} of the
spectral index in the blue regime $n_s>1$ have consistent fits to the cumulative COBE
data as well as to recent ground-base CMB experiments.

In general, {\em exact} inflationary solutions, after an elegant
coordinate transforma\-ti\-on\cite{Sc94}, depend on
the Hubble expansion rate $H$ as a new ``inverse time", and the
regime of inflationary potentials allowing a graceful exit has already been
classified\cite{MS95}.
In our phenomenological approach, the inflationary dynamics is not
prescribed by one's theoretical prejudice. On the contrary, in this
solvable framework, the `graceful exit function' $g(H)$, which
determines the {\em inflaton potential} $U(\phi)$ and  the so--called graceful
exit to the Friedmann cosmos, is {\em reconstructed}\cite{Li95} in order to fit the
data.

\def\rrr{\rule[-3mm]{0mm}{8mm}}
\begin{table}[h]
\caption{\em Eigenvalues of the spectral index $n_s$}\label{table}
\begin{center}
\begin{tabular}{|c|c|c|} \hline
\rrr{\bf Eigenvalue} $n_s$&{\bf Spectrum }&{\bf Potential} $  U(\phi)$ \\
$(-6.57797)$ & (unrealistic) & $\sim\tan^2(\sqrt{\kappa}\phi/4)$ \\
$-3<n_s<1$ & continuous & $\propto \exp (\pm \sqrt{2 \kappa A_0}\; \phi )$ \\
\vdots       & discrete & \vdots \\
1.106   & ($m=5$) & \      \\
1.125   & All CMB data & \      \\
1.153   & ($m=3$) & \      \\
1.194   & COBE data & cf.~Fig.~2 in\cite{SM00} \\
1.247   & ($m=1$) & 
\\
1.49575 & \ & $\sim \tan^2(\sqrt{\kappa}\phi/4)$ \\ \hline
\end{tabular}
\end{center}
\end{table}

Recent astronomical observations by COBE of the cosmic microwave background
(CMB) confirm that the Universe expands rather homogeneously on the large
scale.  Four years CMB of  cumulative observations, have yielded
 $n_{s}=1.2 \pm 0.3$ as
 spectral index, including the quadrupole ($l=2$)
anisotropy.
This experiment has been complemented by ground-based CMB experiments.
Nowaday's best-fit to
{\em all} CMB data\cite{DK00} is a Hubble constant of $H_0=65$ km/(s Mpc) and requires
{\em dark energy} with $\Omega_\Lambda=0.69$ and a
spectral index $n_s=1.12$.

We have applied the $H$--formalism\cite{Sc94} to the rather accurate first order corrections
to the slow-roll approximation of Stewart and Lyth, and transform their nonlinear
equation into the {\em Abel equation} (\ref{Abel}).
The spectral index of our new class\cite{SM00} of solutions can only adopt
the following {\em discrete eigenvalues}
\be
n_s=1+\frac{(2m-1)[4m -1 -4m(2m-1)C]}{2[2m(2m-1)C- 3m +1]^2}
\simeq 1 - \frac{1}{2mC} \; ,
\label{eigenv}
\ee
where, for large $m\ge 1$, we find asymptotically the
scale invariant Harrison--Zel`dovich solution with index $n_s=1$ as
{\em limiting point}. Let us stress that
our discrete spectrum approaches it from the
{\em blue} side, which previously was considered rather difficult to
achieve.
In order to correlate this with observational restrictions, we
display the highest eigenvalues for $n_s$ in Table \ref{table}.
For the new {\em blue} spectrum (\ref{eigenv}) we have
reconstructed for  $m=1$ the potential $U(\phi)$ numerically in Fig. 1 of Ref.\cite{SM00}.

Recently, there have been raised some doubts\cite{DS00} on the applicability of the
second order slow--roll approximation. More recently, however,
 Stewart and Gong \cite{SG01} have explicitly calculated
the spectral index up to {\em second-order corrections}, thereby also corroborating their
previous
next-to leading order  approximation. Then the Abel equation
(\ref{Abel}) turns into a third order nonlinear
differential equation for the energy density with $(2C- \pi^2/6) \epsilon^2 y^3 \epsilon'''$ as
leading term. While our solutions of (\ref{Abel}) are derived from an
{\em even} Taylor expansion in $y=H^2$, the form of the next order corrections suggests that
the additional terms will be {\em odd}
in $y$. Consequently,  our reconstruction on the basis of the Abel equation
is maintained to some approximation, only `orthogonal' terms
 are expected to arise from second order corrections on the percent
level.

\end{document}